\documentclass[9pt,twocolumn,twoside]{osajnl}
\usepackage{comment}
\journal{ol}
\usepackage{txfonts}
\usepackage{amsmath,amssymb}
\usepackage{graphicx}
\setboolean{shortarticle}{true}

\title{Generation of broadband circularly polarized deep-ultraviolet pulses in hollow capillary fibers}
\author[1*]{Athanasios Lekosiotis}
\author[1]{Federico Belli}
\author[1]{Christian Brahms}
\author[1]{John C. Travers}

\affil[1]{School of Engineering and Physical Sciences, Heriot-Watt University, Edinburgh, UK}

\affil[*]{Corresponding author: al104@hw.ac.uk}

\begin{abstract}
We demonstrate an efficient scheme for the generation of broadband, high-energy, circularly polarized femtosecond laser pulses in the deep ultraviolet through seeded degenerate four-wave mixing in stretched gas-filled hollow capillary fibers. Pumping and seeding with circularly polarized 35 fs pulses centered at 400 nm and 800 nm, respectively, we generate idler pulses centered at 266 nm with 27~$\muup$J of energy and over 95\% spectrally averaged ellipticity. Even higher idler energies and broad spectra (27~nm bandwidth) can be obtained at the cost of reduced ellipticity. Our system can be scaled in average power and used in different spectral regions, including the vacuum ultraviolet.
\end{abstract}

\setboolean{displaycopyright}{false}

\begin{document}

\maketitle

High energy and broadband circularly polarized pulses of ultrashort duration in the deep ultraviolet (DUV, 200-300 nm) and vacuum ultraviolet (VUV, 100-200 nm) are a crucial tool for time-resolved spectroscopy of chiral molecules and nanocrystals \cite{boesl_resonance-enhanced_2013,oppermann_ultrafast_2019,besteiro_aluminum_2017}, attosecond pulse generation and metrology \cite{bandrauk_circularly_2017}, circularly polarized high harmonic generation \cite{kfir_helicity-selective_2016} and optically pumped ultrafast magnetism studies \cite{nobusada_photoinduced_2007}. Such pulses are most commonly generated in a two-step process: a nonlinear up-conversion stage in a bulk medium to generate linearly polarized UV pulses followed by a polarization conversion stage using quarter-wave plates or shallow-incidence reflection from metal surfaces \cite{westerveld_production_1985}, reported with low (0.6\%) energy conversion efficiency and few-cycle input pulse requirements \cite{graf_intense_2008}. The order of the generation steps is dictated by the fact that efficient frequency conversion to the DUV in nonlinear crystals requires birefringent phase-matching, precluding direct up-conversion of circularly polarized laser pulses. This approach has two important limitations: low bandwidth and difficulty in reaching the VUV.  Bulk nonlinear crystals have narrow up-conversion bandwidth \cite{baum_generation_2004} and strong material dispersion, limiting the achievable pulse duration. While gases have much lower dispersion, their lower density severely limits the conversion efficiency \cite{beutler_generation_2010}. Material dispersion also prevents achromatic phase retardation in the DUV and VUV, with currently available birefringent phase plates inducing non-uniform phase retardation for different spectral components.

In this work, we extend our previous investigations of efficient frequency up-conversion to the DUV based on seeded four-wave mixing (FWM) in stretched gas-filled hollow capillary fibers (HCF) \cite{belli_strong_nodate}. We demonstrate for the first time, to the best of our knowledge, the direct generation of broadband, ultrashort, circularly polarized DUV pulses (0.95 ellipticity) with pulse energies above $25~\muup$J and a pump-to-idler energy conversion efficiency of $27\%$. The FWM scheme does not require birefringent phase-matching. Therefore, we can pump and seed the process with independently converted circularly polarized pulses at 400~nm and 800~nm, where high-quality achromatic phase retarders are available, and directly generate circularly polarized pulses in the DUV.

The use of hollow waveguides overcomes the aforementioned limitations of bulk media as it does not require a polarization conversion stage for the DUV pulse. It also provides octave-spanning phase-matching bandwidth while dramatically enhancing the light-matter interaction as compared to gas-based techniques in free space. The excellent pump-to-idler frequency up-conversion efficiency provided by FWM schemes has been previously demonstrated with linear polarization in hollow-core photonic crystal fibers (PCFs) \cite{belli_highly_2019} and in HCFs \cite{durfee_ultrabroadband_1997, durfee_intense_1999, misoguti_generation_2001}, reaching a record high of $50\%$ \cite{belli_strong_nodate}. While both PCF and HCF allow for precise control over the generated spectral bandwidth and phase \cite{siqueira_spectral_2016}, stretched HCFs are a fully energy-scalable platform, suitable for pulse compression \cite{nagy_flexible_2008}, FWM interactions \cite{belli_strong_nodate}, and optical soliton dynamics \cite{travers_high-energy_2019}, among other applications. They offer broad transmission free from resonances and birefringence \cite{marcatili_hollow_1964} due to the absence of any micro-structured lattice, which allows access to the VUV and increases the waveguide lifetime.

\begin{figure}[t]
\centering
\includegraphics[width=1\linewidth]{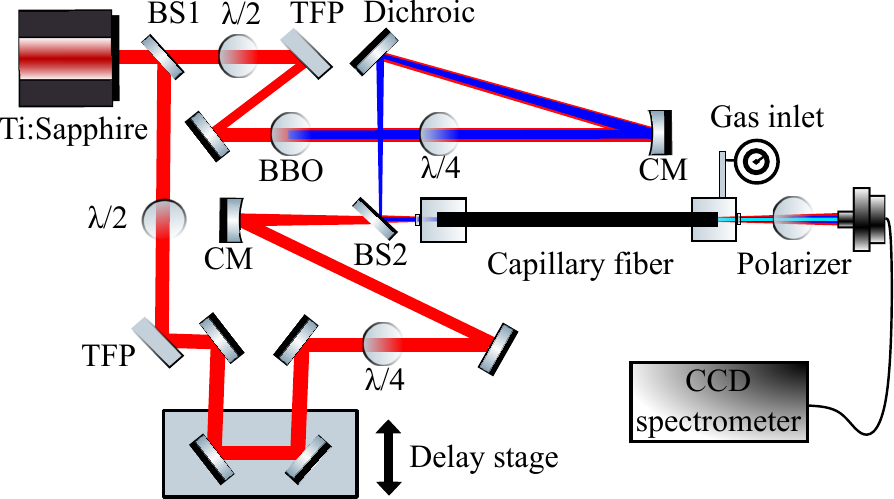}
\caption[FWM experimental setup]%
{Sketch of the experimental apparatus. Beam-splitter (BS1) generates two beam lines, both of which undergo polarization control ($\lambda /2$: half-wave plate, TFP: thin film polarizer). One (pump) undergoes frequency conversion (BBO crystal) and the other (seed) passes through a delay stage. They recombine at BS2 before the input facet of a 1.3 m, $150~\muup$m diameter, stretched capillary fiber, after undergoing focusing by concave mirrors (CM) and conversion to circular polarization with quarter-wave plates ($\lambda /4$). A Rochon polarizer and a CCD spectrometer are used for polarization and spectral energy characterization of the output pulses.}
\label{fig:fig1}
\end{figure}

In our experiments, the output of a 1 kHz, 35 fs FWHM (full width at half maximum), 800 nm Ti:sapphire laser source (Fig.~\ref{fig:fig1}) is split into two replicas (BS1). The first acts as the fundamental seed (signal) while the second is frequency-doubled to 400 nm in a 100~$\muup$m thick $\beta$-Barium borate (BBO) crystal and acts as the degenerate pump. Both pump and seed beam lines are equipped with a motorized half-wave plate (HWP) and a thin-film polarizer (TFP) for power control and conversion to vertical linear polarization with a polarization extinction ratio of $10^{-4}$. A quarter-wave plate (QWP) is placed in each beam line to convert to circular polarization. The pump QWP is slightly tilted off of normal incidence to optimize phase retardation at the central wavelength of the pump. Concave mirrors (CMs) are used as focusing optics to avoid temporal dispersion and chromatic aberration, and a dichroic mirror in the pump line filters out the fundamental 800 nm pulse. The converging beams are combined by a dichroic beam-splitter (BS2) and then propagate to the input of the capillary fiber. A motorized delay stage placed in the seed line enables us to perform delay scans and optimize the temporal overlap of the two pulses.

Stretched hollow capillary fibers provide a substantially improved performance compared with rigid fibers~\cite{nagy_flexible_2008, fan_hollow-core-waveguide_2016, travers_high-energy_2019}. We use our own stretching technique to position a 1.3 m long HCF with a $150~\muup$m core diameter in a sealed gas system that allows for both evacuation and gas fill. Optical access to the waveguide and pressure sealing is provided by two 2~mm thick calcium fluoride (CaF$_{2}$) windows. (CaF$_{2}$ is preferred over other materials as it induces minimum birefringence, even in the UV spectral region \cite{burnett_intrinsic_2001}.) The pulses exiting the capillary are spectrally resolved with a commercial CCD spectrometer equipped with an integrating sphere, which was calibrated as a complete system for spectral power density measurements using a NIST traceable lamp. The wavelength and power density calibration of the spectrometer enables us to measure the pulse energy in different spectral bands simultaneously and directly by integrating over the corresponding spectral region. We additionally cross-calibrate our energy measurement procedure using both a calibrated thermal power meter and calibrated photodiodes.

Spectrally resolved polarization characterization is carried out by measuring the spectral energy density (SED) after transmission through a broadband Rochon linear polarizer as a function of the polarizer rotation angle. An attenuator (not shown in Fig.~\ref{fig:fig1}) is placed at normal incidence before the polarizer to avoid optical damage. The generalized Malus' law states that the transmission of a linear polarizer varies as \cite{azzam_ellipsometry_1987},
\begin{equation}
\label{eqn:ellip}
    T(\Delta\theta)=\frac{\cos^2\Delta\theta+\epsilon^2\sin^2\Delta\theta}{1+\epsilon^2},
\end{equation}
where $\Delta\theta$ is the rotation angle of the polarizer with respect to the angle of maximum transmission and $\epsilon$ is the ellipticity, defined as the ratio of the semi-minor to semi-major axis of the polarization ellipse. Under this definition the ellipticity ranges from 0 for linear polarization to 1 for pure circular polarization. By measuring the transmission of the Rochon polarizer as a function of rotation angle and fitting to Eq.~\ref{eqn:ellip}, we obtain the wavelength-dependent ellipticity $\epsilon (\lambda)$. This can subsequently be weighted with respect to the power spectrum to calculate a spectrally averaged ellipticity ($\bar{\epsilon_{\lambda}}$). Our polarization characterization routine enables us to measure the ellipticity of multiple spectral bands simultaneously with a standard deviation error of $\pm0.005$, corresponding to an average SED change of $\pm0.25\%$ relative to its maximum value due to fluctuations during data acquisition.

\begin{figure}[b]
\centering
\includegraphics[width=1\linewidth]{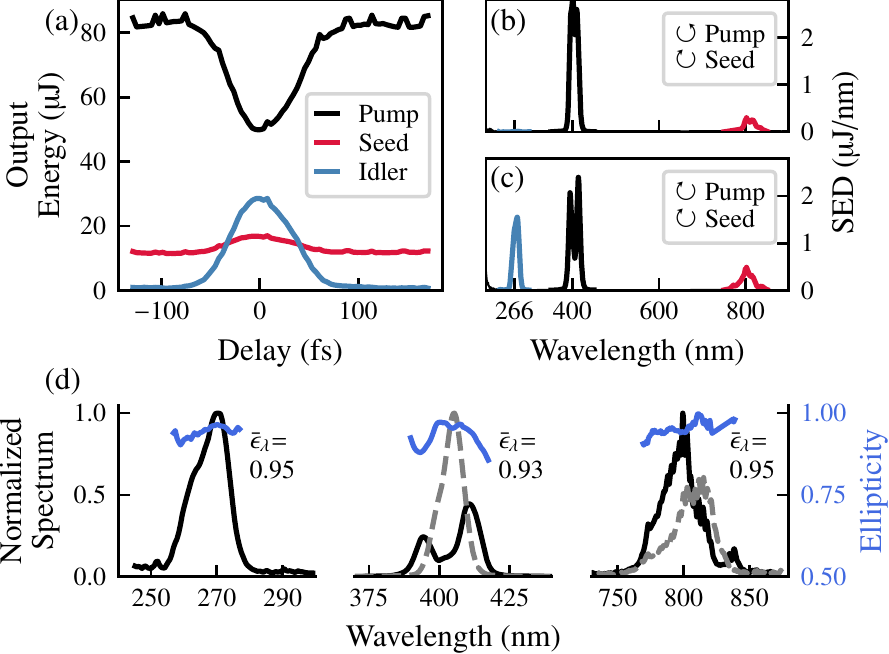}
\caption[Circularly-polarized FWM]%
{(a) Idler UV (blue) generation relative to time delay between pump (black) and seed (red) for co-polarized inputs. Pump and seed coupled energies are $100~\muup$J and $27~\muup$J, respectively. At zero delay, a DUV idler with $27~\muup$J energy is generated with $27~\%$ coupled pump to idler conversion efficiency. SED measurements for cross-polarized (b) and co-polarized (c) input configuration at optimal temporal overlap, showing zero and maximum idler generation (blue), respectively, based on spin conservation requirement. (d) Normalized spectra of (a) and (c) and the corresponding polarization profiles (solid blue). Spectrally-weighted ellipticities ($\bar{\epsilon_{\lambda}}$) are 0.95 for the idler and seed and 0.93 for the pump. Dashed grey lines indicate the input spectra.}
\label{fig:fig2}
\end{figure}

We investigate the energy and polarization transfer efficiency from the input fields to the DUV idler in both the weak-seeding (coupled seed energy $<60~\muup$J) and strong-seeding (coupled seed energy $>100~\muup$J) regimes. In the weak-seeding regime ($27~\muup$J), we fill the HCF with helium at $\sim\!1.8$ bar (which maximizes the DUV energy) and pump the FWM process with $100~\muup$J. After scanning the delay between the two circularly polarized input pulses (Fig.~\ref{fig:fig2}a) to find the optimum, a broadband circularly polarized UV idler centered at 266~nm is generated with a pulse energy of 27~$\muup$J, corresponding to a $27\%$ pump-to-idler conversion efficiency (CE, defined as the energy ratio of the generated idler to the coupled pump). 

The conservation of the spin angular momentum requires the two input pulses to have the same state of polarization (SOP) for FWM to occur \cite{lin_vector_2004}. For the case of circular polarization, this corresponds to both input pulses being left or right circularly polarized. If the input pulses have opposite-circular SOP, no FWM takes place and idler generation is inhibited (Fig.~\ref{fig:fig2}b) despite spatio-temporal overlap of the input pulses. The ellipticity of the output UV pulse shown in Fig.~\ref{fig:fig2} is 0.95, very close to purely circularly polarized, with a flat, uniform polarization profile (blue line in Fig.~\ref{fig:fig2}d) across its spectrum (black line). As shown by the difference between the black and grey dashed lines in Fig.~\ref{fig:fig2}d, the frequency conversion process reshapes the spectrum of both the pump and seed pulses due to Kerr-induced nonlinear propagation phenomena such as self- and cross-phase modulation (SPM, XPM) \cite{stolen_self-phase-modulation_1978, islam_cross-phase_1987} in addition to strong pump energy depletion ($\sim40$\%, Fig.~\ref{fig:fig2}a). Despite these spectral shifts, we measure pump and seed output ellipticities of 0.93 and 0.95, a negligible change from their input values of 0.945 and 0.95. 


\begin{figure*}[ht]
\centering
\includegraphics[width=1\linewidth]{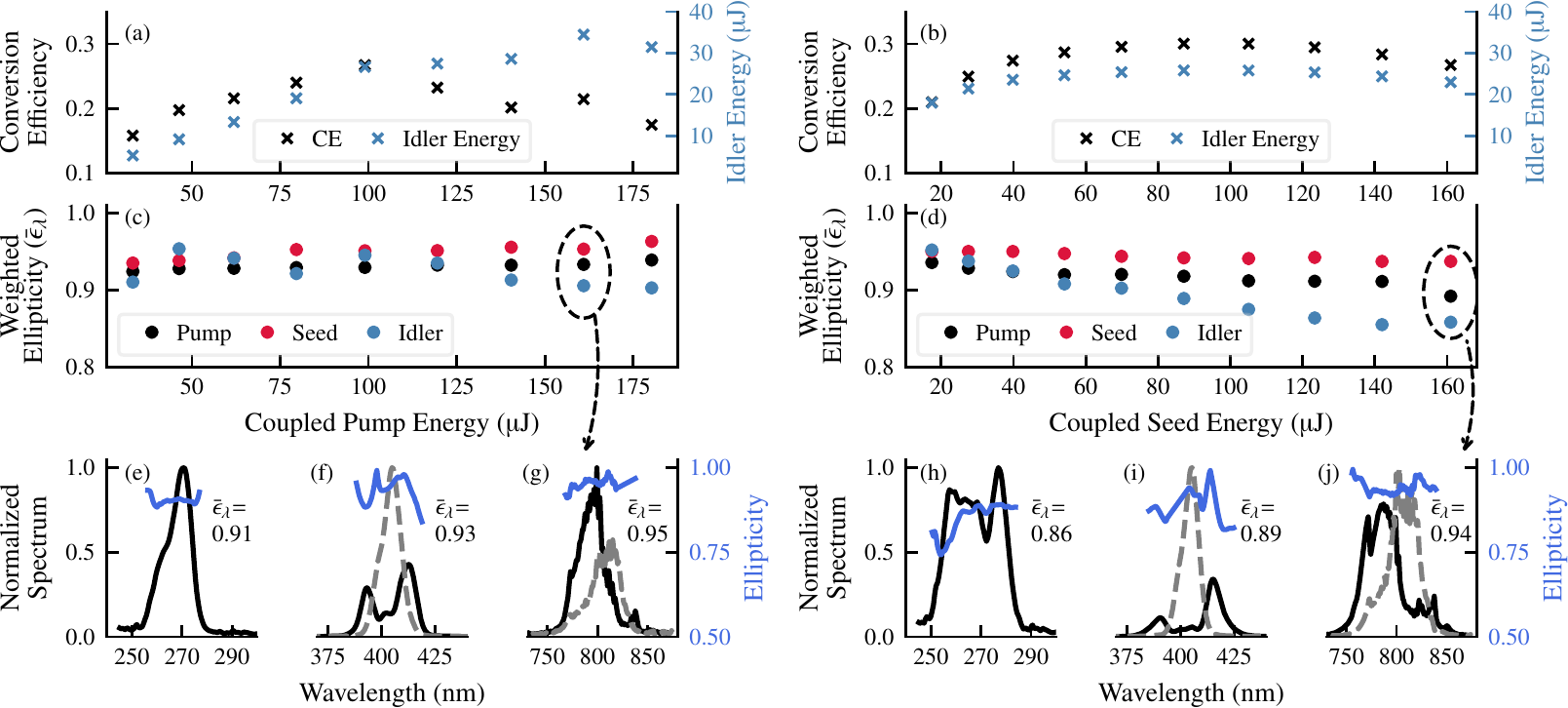}
\caption[Polarization in high-energy FWM]%
{Pump (left-hand side) and seed (right-hand side) energy scans with seed and pump coupled energy fixed at $27~\muup$J and $80~\muup$J, respectively. Pump-to-idler energy conversion efficiency (black) and output idler pulse energy (blue) relative to pump (a) and seed (b) coupled energies. Ellipticity measurements for the pump (black), seed (red) and idler (blue) relative to pump (c) and seed (d) coupled energies. Output spectra and polarization profiles (e-j) of pump, seed, idler for pump and seed energy yielding maximum idler energy (e-g) and broad idler spectra (h-j).}
\label{fig:fig3}
\end{figure*}

The driving energies shown in Fig.~\ref{fig:fig2} are chosen to simultaneously optimize the conversion efficiency and polarization purity of the idler pulse. Increasing the pump and seed energies increases the strength of nonlinear effects. For imperfectly circularly polarized light ($\epsilon\ne1.0$), it has previously been shown that nonlinear phenomena can modify the polarization state by inducing birefringent nonlinear phase shifts \cite{qiang_lin_vector_2004}. This results in non-uniform polarization profiles and decreased ellipticities. This polarization evolution affects not only the two input pulses but also the generated idler. In Fig.~\ref{fig:fig3} we present the response of our FWM system when varying the energy of either the pump (left column) or the seed (right column) while keeping the other fixed at $27~\muup$J (seed) and $80~\muup$J (pump).

In the pump energy scan, a maximum CE of 27\% is obtained for $100~\muup$J yielding $27~\muup$J in the DUV (the conditions shown in Fig.~\ref{fig:fig2}), whilst the idler energy continues to increase beyond $100~\muup$J pumping, reaching $34~\muup$J for $160~\muup$J pump energy. In the seed energy scan, CE and idler energy reach 30\% and $34~\muup$J, respectively, and slowly decrease for seed energies above $120~\muup$J. The ellipticities of pump and seed increase with the pump energy, from 0.925-0.94 and from 0.935-0.965, respectively (Fig.~\ref{fig:fig3}c). The idler ellipticity remains relatively constant around 0.95 and decreases to 0.91 beyond $100~\muup$J pump energy, following the CE trend. In the seed energy scan (Fig.~\ref{fig:fig3}d), both driving pulses exhibit only minor changes in their ellipticity, with the pump ellipticity constant at about 0.93 and the seed at 0.94-0.95. The idler ellipticity, on the other hand, decreases from 0.95 at low seed energies to 0.86 at $160~\muup$J.

Figures \ref{fig:fig3}e-j show the pump, seed and idler spectra and their corresponding polarization distributions at specific energy points of interest. From the pump scan, we select the pump energy yielding the highest idler energy ($34~\muup$J in the DUV, obtained for $160~\muup$J pump energy) and from the seed scan the maximum seed energy. Strong spectral modifications of the pump (depletion) and seed (shifts associated with amplification and pump-induced XPM) are evident in both cases, as compared to their input profiles (grey dashed). The spectra associated with the strong-seeding regime (Fig.~\ref{fig:fig3}h-j) show a more dramatic enhancement of the nonlinear effects. The presence of the strong seed affects the polarization uniformity of both the pump (Fig.~\ref{fig:fig3}i) and the seed (Fig.~\ref{fig:fig3}j) as compared to the ones for the weak-seeding regime, shown in Fig.~\ref{fig:fig2}d, resulting in a decreased ellipticity (0.89) for the pump. The idler generated in the strong-pump regime (Fig.~\ref{fig:fig3}e) maintains a flat polarization profile with 0.91 ellipticity. In the strong-seeding case (Fig.~\ref{fig:fig3}h), a broadband (27~nm FHWM) DUV idler with 0.86 ellipticity is generated with $\sim\!30\%$ conversion efficiency and pulse energy above $20~\muup$J. Although direct temporal characterization is challenging, this bandwidth is sufficient to support a 4~fs pulse duration. 

The slight idler ellipticity decrease mainly observed in the strong-seeding regime cannot be solely attributed to a single phenomenon due to the coupled nature of nonlinear effects taking place in FWM. However, the nonlinear polarization evolution induced by each pulse upon itself can be isolated by launching the input beams separately in 1.8 bar of helium and measuring the output polarization. By evacuating the fibre (to avoid nonlinearity) we can also estimate the polarization of the input pulses from measurements at the fiber output, due to the negligible birefringence of the HCF and the CaF$_{2}$ windows. 

\begin{figure}[t]
\centering
\includegraphics[width=1\linewidth]{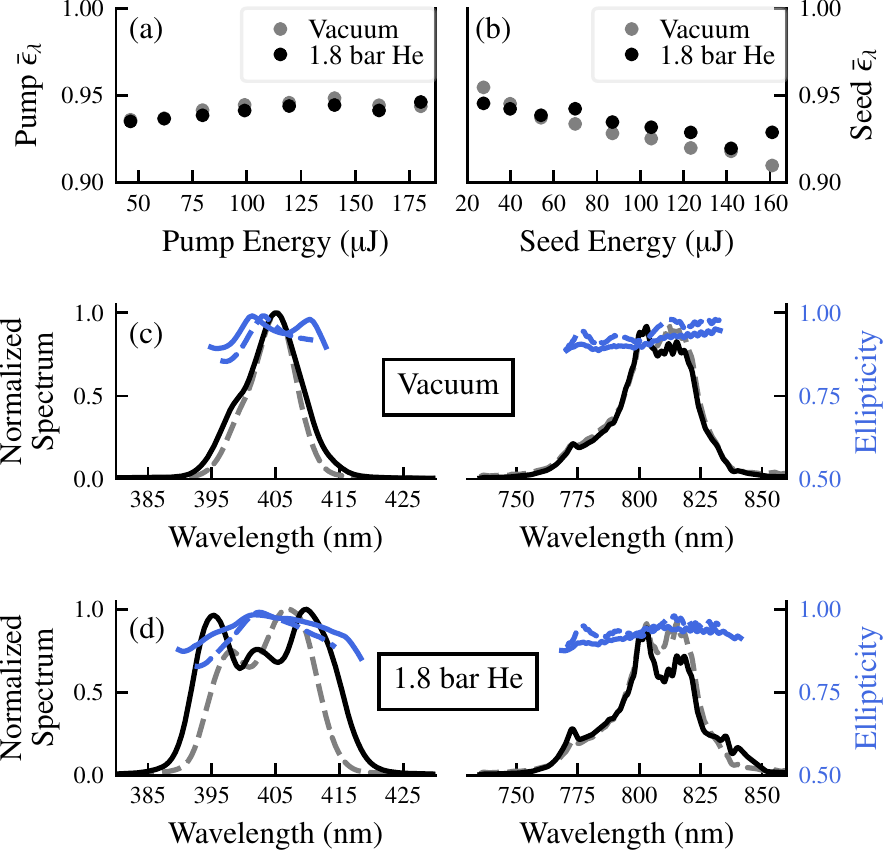}
\caption[Input Polarization]%
{(a) and (b) Polarization measurements for the pump and seed relative to their driving energies in vacuum (gray circles) and in 1.8 bar of helium (black circles). High-energy (>$125~\muup$J for the pump, >$100~\muup$J for the seed, solid black lines) and low-energy (<$85~\muup$J for the pump, <$60~\muup$J for the seed, dashed gray lines) averaged spectra and their averaged polarization profiles (solid blue and dashed blue lines, respectively) as measured in vacuum (c) and in 1.8 bar of helium (d).}
\label{fig:fig4}
\end{figure}

In Fig.~\ref{fig:fig4}(a, b) we compare the polarization evolution of the input pulses (in vacuum, gray circles) and under nonlinear conditions (1.8~bar He, black circles) for the pump and seed by themselves. It can be observed that both the pump and the seed exhibit negligible nonlinear polarization evolution by themselves compared to the input pulse measurements. Indeed, the pump is measured to have an ellipticity of 0.94 in both cases, and the seed exhibits a slightly decreasing ellipticity from 0.95 at low energies to about 0.92 at higher energies, similar to the input pulse trend. In Fig.~\ref{fig:fig4}(c, d) we show the averaged pump and seed spectra for both high (solid black) and low energy (dashed gray) with the corresponding averaged ellipticities (solid and dashed blue, respectively) for the two pressure sets. In the vacuum set (Fig.~\ref{fig:fig4}c), the spectral broadening of the pump indicates the presence of nonlinearity along the beam path (possibly at the CaF$_{2}$ windows due to the high intensity of the converging beams), which may explain the slight change of input seed ellipticity with varying energy (Fig.~\ref{fig:fig4}b). Finally, in the 1.8~bar helium set (Fig. \ref{fig:fig4}d), the high-energy pulses maintain both polarization uniformity and high ellipticity values, despite the broadened spectral profiles.

In conclusion, we have demonstrated an efficient method for the generation of ultrashort DUV pulses with a uniform spectral distribution of circular (0.95 ellipticity) polarization, high pulse energy (above 25~$\muup$J) and 27\% pump-to-idler conversion efficiency, via weakly seeded FWM in stretched HCFs. Despite the slight idler ellipticity decrease (0.86) in the strong-seeding regime due to nonlinear propagation effects, we have also generated broadband (27~nm FWHM) circularly polarized DUV spectra supporting 4 fs pulse durations. The system described can be readily adapted to other spectral regions, including the VUV, by proper selection of the driving frequencies, and scaled in average power by increasing the repetition rate.


\section*{Disclosures}
The authors declare no conflicts of interest.

\bibliography{MyLibrary}
\bibliographyfullrefs{MyLibrary}
\end{document}